\documentclass[conference]{IEEEtran}
\usepackage[utf8]{inputenc}
\usepackage{clrs+} 
\usepackage{amsmath,amssymb,amsfonts}
\usepackage{algorithm}
\usepackage[noend]{algpseudocode}
\usepackage{graphicx}
\usepackage{subcaption}
\usepackage{multirow}
\usepackage{booktabs}
\usepackage{colortbl}

\newcommand{\mat}[1]{\mathbf{#1}}

\newcommand{\nnz}{\textit{nnz}}

\newcommand{\spkadd}{\texttt{SpKAdd~}}
\newcommand{\coladd}{\texttt{ColAdd~}}
\newcommand{\erdosrenyi}{Erd\H os-R\'{e}nyi}

\newcommand{\mA}{\mathbf{A}} 
\newcommand{\mB}{\mathbf{B}}

\title{\huge{Parallel Algorithms for \\Adding a Collection of Sparse Matrices}}
\author{\IEEEauthorblockN{
    Md Taufique Hussain\IEEEauthorrefmark{1},
    Guttu Sai Abhishek\IEEEauthorrefmark{2}
    Aydin Bulu\c{c}\IEEEauthorrefmark{3}
    and
    Ariful Azad\IEEEauthorrefmark{4} }  
  \IEEEauthorblockA{\IEEEauthorrefmark{1}
    Indiana University, Bloomington, IN, USA (mth@indiana.edu)}
    \IEEEauthorblockA{\IEEEauthorrefmark{2}
    IIT Bombay, India (180050036@iitb.ac.in)}
  \IEEEauthorblockA{\IEEEauthorrefmark{3}
    Lawrence Berkeley National Laboratory, Berkeley, CA, USA (abuluc@lbl.gov)}
  \IEEEauthorblockA{\IEEEauthorrefmark{4}
    Indiana University, Bloomington, IN, USA (azad@iu.edu)}
}  
\date{}

\begin{document}

\maketitle

\begin{abstract}
    We develop a family of parallel algorithms for the SpKAdd operation that adds a collection of k sparse matrices. 
    SpKAdd is a much needed operation in many applications including distributed memory sparse matrix-matrix multiplication (SpGEMM), streaming accumulations of graphs, and algorithmic sparsification of the gradient updates in deep learning.
    While adding two sparse matrices is a common operation in Matlab, Python, Intel MKL, and various GraphBLAS libraries, these implementations do not perform well when adding a large collection of sparse matrices.
    We develop a series of algorithms using tree merging, heap, sparse accumulator, hash table, and sliding hash table data structures. Among them, hash-based algorithms attain the theoretical lower bounds both on the computational and I/O complexities and perform the best in practice. The newly-developed hash SpKAdd makes the computation of a distributed-memory SpGEMM algorithm at least 2x faster than that the previous state-of-the-art algorithms.    
\end{abstract}

\pagenumbering{roman}
\section{Introduction}
Sparse matrices 
power many modern scientific codes, including those that arise in simulations, data analytics, and machine learning. An overlooked operation that is common across many applications is the summation (or reduction) of a collection of sparse matrices. In the deep learning community, this is called the \emph{sparse allreduce} problem~\cite{zhao2014kylix,renggli2019sparcml} that arises due to algorithmic sparsification of the gradient updates for reduced communication. While most papers describe the problem as the reduction of sparse vectors, each holding gradient updates from one processor, in practice the computation becomes the reduction of a collection of sparse matrices due to mini-batching commonly used in gradient-based deep learning training. The existing work on this sparse allreduce problem is exclusively on the communication and distributed-memory aspects, leaving the actual in-node reduction itself understudied.   

The use case that primarily motivated our work is the distributed-memory multiplication of two sparse matrices (SpGEMM). In virtually all algorithms for the distributed memory SpGEMM, there is a step where each processor has to sum intermediate products of the form $\sum_{i=1}^k \mA_i \mB_i$. We call this operation {\tt SpKAdd}. While local SpGEMM algorithms have seen significant advances in their performance within the last 15 years~\cite{Nagasaka2018}, {\tt SpKAdd} has not received any attention, leaving it a potential bottleneck. This is especially true when we try to achieve strong scaling or when there is overdecomposition of the problem for ease of load balancing: the more processes perform distributed SpGEMM, the more computation shifts from local SpGEMMs to {\tt SpKAdd}. This seemingly counterintuitive fact is easy to verify when considering the extreme case where each process owns a single $1{\times} 1$ matrix and there are $n^2$ processors for multiplying two $n{\times} n$ sparse matrices: {\tt SpKAdd} does half the flops (i.e., all the additions) needed for the distributed SpGEMM. 
Furthermore, communication-avoiding SpGEMM algorithms~\cite{lazzaro2017increasing, batchedSpGEMMIPDPS21} utilize {\tt SpKAdd} at two different phases: one within each 2D grid of the overall 3D process grid and another when reducing results across different 2D grids.

Finally, {\tt SpKAdd} is needed for assembling local finite-element matrices into a global matrix~\cite{fu2014architecting}. This problem has traditionally been labeled as one that presents few opportunities for parallelism. In this work, we show this not to be true. In fact, {\tt SpKAdd} has plenty of parallelism. However, existing implementations use suboptimal data structures for the accumulation, hurting their performance. In this paper, we systematically evaluate various different data structures and methods for {\tt SpKAdd} in shared memory. 

The main contribution of this paper is to develop efficient parallel \texttt{SpKAdd} algorithms that are work-efficient, use spatial locality when accessing memory, and do not require thread synchronization.
We achieve these goals by utilizing heap, sparse accumulator (SPA)~\cite{gilbert1992sparse}, hash table data structures. 
We designed a sliding hash algorithm 
after carefully considering (a) the data access patterns for different data structures, (b) matrix shapes and sparsity patterns, and (c) hardware features such as the cache size.
We theoretically and experimentally demonstrate that hash and sliding hash based \spkadd algorithms run the fastest for input matrices with various shapes and sparsity patterns on three different shared-memory platforms. 
We demonstrate the impact of fast \spkadd algorithms by plugging the hash algorithm in a distributed SpGEMM implementation in the CombBLAS library~\cite{combblas2}.  
When running SpGEMM on 2K nodes of a Cray XC40 supercomputer, the hash-based \spkadd made the computation of the distributed SpGEMM $2\times$ faster than the previous implementation. 

\section{Algorithmic landscape of the \spkadd operation}
\subsection{The problem}
We consider the operation \spkadd that adds $k$ sparse matrices $\mA_1, \mA_2,...,\mA_k$ and returns the summed (potentially sparse) matrix $\mB {=} \sum_{i=1}^k \mA_i$, where $\mA_i {\in} \mathbb{R}^{m \times n}$ and $\mB {\in} \mathbb{R}^{m \times n}$. 
We assume that all input matrices $\mA_1, \mA_2,...,\mA_k$ are available in memory before performing the \texttt{SpKAdd} operation.
Let $\nnz(\mA_i)$ denote the number of nonzeros in $\mA_i$. 
Then, $\nnz(\mB) \leq \sum_{i=1}^k \nnz(\mA_i)$.
The compression factor \id{cf} of an \spkadd operation is defined as $\sum_{i=1}^k \nnz(\mA_i)/\nnz(\mB)$, where $\id{cf}{\geq} 1$.
When $k{=}2$, the operation simply adds two sparse matrices. In this special case of $k{=}2$, \texttt{SpKAdd} is equivalent to \texttt{mkl\_sparse\_d\_add} in MKL, the ``+" operator in Matlab and Python (with scipy sparse matrices as operands). 
While adding two sparse matrices sequentially or in parallel is a widely-implemented operation in any sparse matrix library, \texttt{SpKAdd} is rarely implemented in most libraries. 
One can repeatedly add sparse matrices in pairs to add $k$ sparse matrices.
However, we will analytically and experimentally demonstrate that the pairwise addition could be significantly slower than a unified \texttt{SpKAdd} operation when $k$ is large.

Throughout the paper, we assume that input and output matrices are represented in the compressed sparse column (CSC) format that stores nonzero entries column by column. However, all algorithms discussed in this paper are equally applicable to compressed sparse row (CSR), coordinate (COO), doubly compressed, and other sparse matrix formats~\cite{langr2015evaluation}. 

\setlength{\textfloatsep}{5pt}
\begin{algorithm}[!t]
\caption{SpKAdd using 2-way addition of columns.}
\textbf{Input:} A list of $k$ sparse matrices $\mA_1,...,\mA_k$. {\bf Output:} $\mat{B}$. All matrices are in $\mathbb{R}^{m \times n}$
\begin{algorithmic}[1]

\Procedure{SpKAdd2Way}{$\mA_1,...,\mA_k$}
\State $\mB\gets \mA_1$
\For{$i \gets 2$ to $k$}
     \For{$j \gets 1$ to $n$}
     \State $\mB(:,j) \gets$ ColAdd ($\mB(:,j), \mA_i(:,j)$)  
\EndFor 
\EndFor

\State \Return $\mB$
\EndProcedure
\end{algorithmic}
\label{algo:twoway}
\end{algorithm}
\subsection{\spkadd using 2-way additions}
\vspace{-3pt}
At first, we describe \spkadd algorithms that rely on 2-way addition of a pair of matrices. We have implemented parallel 2-way additions in our library, but an implementation from MKL, Matlab, and GraphBLAS~\cite{bulucc2017design, davis2019algorithm} can be used. 
\subsubsection{\spkadd using 2-way incremental additions}
The simplest approach to implement \spkadd is to incrementally add $k$ sparse matrices in pairs.
Algorithm~\ref{algo:twoway} provides a skeleton of \spkadd implemented with 2-way incremental additions.  
Here, the $i$th iteration of the outer loop adds the current partial result with the next input matrix $\mA_i$. 
Since matrices are stored in the CSC format, the $j$th columns of $\mA_i$ and $\mB$ can be added independently as denoted by the \coladd operation in line 5 of Algorithm~\ref{algo:twoway}. 
Here, the $j$th columns of a matrix is an array of \texttt{(rowid, val)} tuples where \texttt{rowid} denotes the row index of a nonzero entry. 
If columns are sorted in the ascending order of row index, the \coladd function simply merges two lists of tuples, which is similar to the merging operation of the merge sort algorithm.
Fig.~\ref{fig:AllAlgorithms}(b) shows an example of 2-way incremental addition for the $j$th columns of four matrices shown in Fig.~\ref{fig:AllAlgorithms}(a). 


{\bf Computational complexity.} 
First, consider the cost of adding two matrices $\mA_1$ and $\mA_2$. 
If each columns are sorted based on their row indices, a linear merging algorithm has $O(\nnz(\mA_1)+\nnz(\mA_2))$ complexity. 
Even though faster merging algorithm exists for imbalanced lists ~\cite{brown1979fast}, they often are too complicated for parallel implementations. 
In the worst case where the matrices have no common row indices, $\nnz(\mA_1+\mA_2) = \nnz(\mA_1)+\nnz(\mA_2)$. This is especially true when nonzeros in the input matrices are uniformly distributed such as in \erdosrenyi~(ER) random matrices.
In this worst-case model, the $i$th iteration of the 2-way incremental algorithm has the cost of $\sum_{l=1}^i{\nnz(\mA_l)}$ and the total cost of the algorithm is
$\sum_{i=2}^k\sum_{l=1}^i{\nnz(\mA_l)}$.
If all input are ER matrices with $d$ nonzeros per column, the overall computational cost is $\sum_{i=2}^k\sum_{l=1}^i{nd} = O(k^2nd)$.

{\bf I/O complexity.} 
In this paper, the I/O complexity measures the amount of data transferred to and from the main memory. 
If we assume that the output of every 2-way addition is stored in the main memory,  the I/O complexity of the 2-way incremental algorithm matches the computational complexity discussed above. That means $O(k^2nd)$ data is accessed from the main memory for ER matrices.

\subsubsection{\spkadd using 2-way tree additions}
\begin{figure*}
    \centering
    \includegraphics[width=\linewidth]{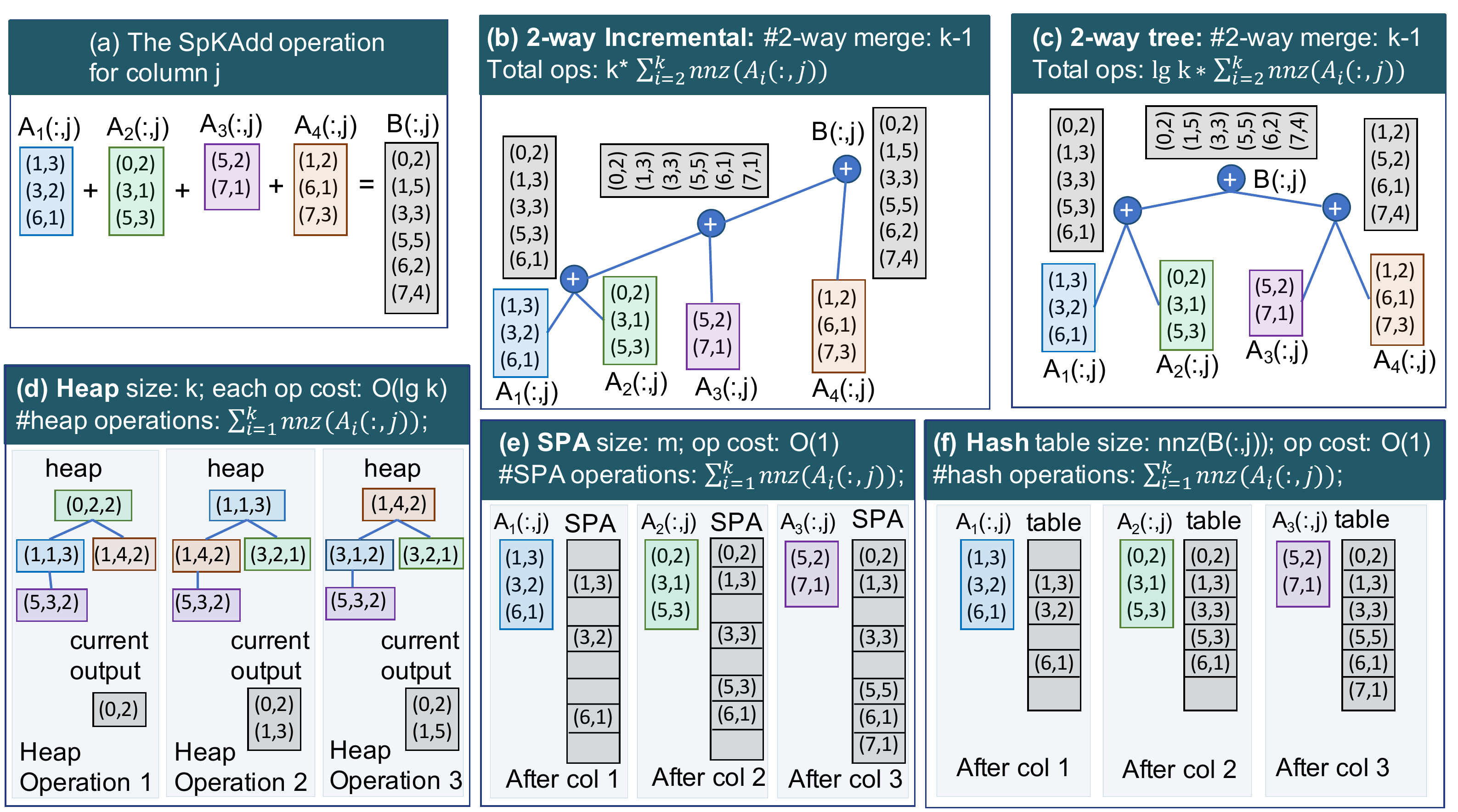}
    \vspace{-12pt}
    \caption{
    Data structure and algorithmic choices when adding the $j$th columns of four input matrices. (a) Input and  output columns, where each column stores (rowid, value) tuples. Different colors are used to show different input columns. (b) 2-way incremental addition. (c) 2-way tree addition where inputs are at the leaves of a binary tree. (d) k-way addition using a min-heap of size $k=4$. The heap stores (rowid, matrixid, value) tuples and holds at most one entry from each input column.  The entry with the minimum row index is at the root of the tree representing the min-heap.  Only the first three heap operations (extracting the minimum entry and inserting another entry) are shown. We also show the current status of the output column. (e) k-way addition using a SPA of size $m=8$. Any entry from the $i$th row of an input matrix such as $\mA_1(i,j)$ is added to the $i$th entry of the SPA. (f) k-way addition using a hash table of size $\mB(:,j)=6$. For simplicity of depiction, we assume that the hash function maps row indices to hash table in the ascending order of row indices in $\mB(:,j)$ (e.g., $\mB(5,j)$) is the 4th entry of the output and mapped to the 4th position of the hash table). \vspace{-5pt}}
    \label{fig:AllAlgorithms}
\end{figure*}

The 2-way incremental addition is expensive both in terms of the computational complexity and data movements. 
One reason of this inefficiency is the imbalanced nature of the binary addition tree shown in Fig.~\ref{fig:AllAlgorithms}(b) where the height of the tree is $k-1$.
To address this problem, we consider a balanced binary tree as shown in Fig.~\ref{fig:AllAlgorithms}(c) where the height of the tree is $\lg k$.
In the 2-way tree addition, input matrices are added in pairs in the leaves of the tree before the resultant matrices are added at internal nodes. 
Thus, the computational complexity of all additions at a given level of the binary tree is $O(\sum_{i=1}^k{\nnz(\mA_i)})$, which results in the overall complexity of $O(\lg k \sum_{i=1}^k{\nnz(\mA_i)})$.
Assuming that the intermediate matrices are stored in memory, the amount of data movement is also $O(\lg k \sum_{i=1}^k{\nnz(\mA_i)})$.
For ER input matrices, the computational and data movement complexity become $O(ndk\lg k)$.
Hence, for any large value of $k$, 2-way tree addition is faster than 2-way incremental addition both in theory and in practice.
This performance improvement comes for free as we can still use any off-the-shelf function to add a pair of matrices.  

\vspace{-5pt}

\begin{algorithm}[!t]
\caption{SpKAdd using k-way addition of columns.}
\textbf{Input:} A list of $k$ sparse matrices $\mA_1,...,\mA_k$; \textbf{Output:} $\mat{B}$. All matrices are in $\mathbb{R}^{m \times n}$
\begin{algorithmic}[1]

\Procedure{SpKAdd}{$\mA_1,...,\mA_k$}
\For{$j \gets 1$ to $n$}
     \State $\mB(:,j) \gets$ ColAddKWay ($\mA_1(:,j),...,\mA_k(:,j)$)  
     \Comment{HeapAdd, SPAAdd, HashAdd, etc.}
\EndFor
\State \Return $\mB$
\EndProcedure
\end{algorithmic}
\label{algo:generickway}
\end{algorithm}

\begin{algorithm}[!h]
\caption{SpKAdd using a min-heap. Only the addition of $j$th columns is shown.
The $j$th column $\mA(:,j)$ is stored as a list of tuples $(r,v)$ where $r$ and $v$ denote row index and value. 
}

\begin{algorithmic}[1]
\Procedure{HeapAdd}{$\mA_1(:,j),...,\mA_k(:,j)$}
    \State $\id{Heap}  \gets \phi$ 
    \For{$i \gets 1$ to $k$}
    \State  $(r,v) \gets $ the nonzero entry with the smallest row index in $\mA_i(:,j)$
    \State  \Call{Insert}{\id{Heap}, $(r,i,v)$}
    \EndFor
     
     \While{\id{Heap} $\neq \phi$}
        \State $(r,i,v) \gets $ \Call{ExtractMin}{\id{Heap}}
        \If{$\mB(r,j)$ exists}
            \State $\mB(r,j) \gets \mB(r,j) + v$
        \Else \Comment{append at the end of the output}
            \State $\mB(:,j) \gets \mB(:,j) \cup (r,v)$ 
        \EndIf
        \If{$\mA_i$ still has unused entries} 
        \State $(r_{next},v_{next}) \gets $ the next element from $\mA_i$
        \State  \Call{Insert}{\id{Heap}, $(r_{next},i,v_{next})$}
        \EndIf
        
     \EndWhile
\State \Return $\mB(:,j)$
\EndProcedure
\end{algorithmic}
\label{algo:heap}
\end{algorithm}

\begin{algorithm}[!h]
\caption{SpKAdd using a SPA. Only the addition of $j$th columns is shown.
The $j$th column $\mA(:,j)$ is stored as a list of tuples $(r,v)$ where $r$ and $v$ denote row index and value. 
}

\begin{algorithmic}[1]
\Procedure{SPAAdd}{$\mA_1(:,j),...,\mA_k(:,j)$}
    
    \State $\id{SPA}  \gets $ a dense array of length $m$ 
    \State $\id{idx}  \gets \phi$  \Comment{a list of valid indices in \id{SPA} }
    \For{$i \gets 1$ to $k$}
        \For{$(r,v) \in \mA_i(:,j)$}
            \If{$r \in \id{idx}$ }
                $\id{SPA}[r]\gets \id{SPA}[r] + v$
            \Else
                \ \ $\id{SPA}[r]\gets  v$;\  $\id{idx} \gets \id{idx}\cup r$ 
            \EndIf
        \EndFor
    \EndFor
\State \Call{Sort}{\id{idx}} \Comment{if sorted output is desired}
\For{$r \in \id{idx}$ }
    \State append  $(r,\id{SPA}[r])$ into $\mB(:,j)$
\EndFor
\State \Return $\mB(:,j)$
\EndProcedure
\end{algorithmic}
\label{algo:SPA}
\end{algorithm}

\begin{algorithm}[!h]
\caption{SpKAdd using a hash table. Only the addition of $j$th columns is shown.
The $j$th column $\mA(:,j)$ is stored as a list of tuples $(r,v)$ where $r$ and $v$ denote row index and value. 
}

\begin{algorithmic}[1]
\Procedure{HashAdd}{$\mA_1(:,j),...,\mA_k(:,j)$}
    
    \State $\id{HT}  \gets $ an array whose size is a power of two and greater than $\nnz(\mB(:,j))$; initialized with $(-1,0)$
    \For{$i \gets 1$ to $k$}
        \For{$(r,v) \in \mA_i(:,j)$}
            \State $h\gets \Call{Hash}{r}$ \Comment{hash index from row index}
            \While {true}
            \If{$\id{HT[h]}$ is empty}
            \State $\id{HT}[h]\gets  (r,v)$; {\bf break}
            \ElsIf{$r$ is found at $\id{HT[h]}$ }
            \State add $v$ to the value at  $\id{HT}[h]$; {\bf break}
            \Else \Comment{hash conflict resolution}
                \State $h\gets $ next hash index
            \EndIf
        \EndWhile
        \EndFor
    \EndFor
\For{every valid hash index $h \in \id{HT}$ }
    \State append  $\id{HT}[h]$ into $\mB(:,j)$
\EndFor
\State \Call{Sort}{$\mB(:,j)$} \Comment{if sorted output is desired}
\State \Return $\mB(:,j)$
\EndProcedure
\end{algorithmic}
\label{algo:hash}
\end{algorithm}

\subsection{\spkadd using k-way additions}
\vspace{-3pt}
Since repeated use of 2-way additions leads to an inefficient SpKAdd operation, we developed several algorithms that use k-way additions. 
Algorithm~\ref{algo:generickway} provides a template for such an algorithm where $j$th columns of all input matrices are added to form the $j$th column of the output. 
Line 3 of Algorithm~\ref{algo:generickway} calls the k-way addition of columns for which we considered several data structures including heaps, sparse accumulators, and hash tables. 

\subsubsection{HeapSpKAdd: k-way additions using a heap}
Algorithm~\ref{algo:heap} shows the HeapAdd operation that uses a min-heap to add $j$th columns of all input matrices. Here, \id{Heap} represents a binary tree storing $k$ tuples with each $(r,i,v)$ tuple represents $v=\mA_i(r,j)$, i.e., $v$ is stored in the $r$th row and $j$th column of $\mA_i$. 
Each tuple in the heap has a unique matrix index (the middle entry of the tuple), meaning that at most one tuple in the heap is from an input matrix.  
The min-heap uses row indices as the key and stores the tuple with the minimum row index at the root. 
Hence, extracting a tuple with the minimum row index takes $O(1)$ time, but inserting a new entry to the heap takes $O(\lg k)$ time.
See Fig.~\ref{fig:AllAlgorithms}(d) for an example. 

Lines 3-5 of Algorithm~\ref{algo:heap} initialize the heap with the smallest row indices from all input columns. 
Each iteration of the loop at line 6 extracts the minimum entry $(r,i,v)$ (based only on the row index) from the heap. If this is the first time we extract an entry from row $r$, we append $(r,v)$ at the the end of the output.
If there is already an entry in the output column at row $r$, we simply add $v$ to this row index. Since the output column is formed in the ascending order of row indices, the existence of $\mB(r,j)$ in line 8 can be checked in $O(1)$ time. If the minimum entry from the heap came from $\mA_i$, we insert the next entry from $\mA_i$ into the heap (line 12-14).
Here, the heap algorithm assumes that all columns of input matrices are sorted in the ascending order of row indices.

{\bf Complexity:} Inserting a tuple to the heap takes $O(\lg k)$ time. Since every nonzero entry from the input must be inserted to the heap for processing, the total computational complexity of the heap based algorithm is $O(\lg k \sum_{i=1}^k{\nnz(\mA_i)})$, which is equivalent to the 2-way tree algorithm algorithm.
For I/O complexity, we assume that 
the heap (of size $O(k)$) fits in a thread-private cache and HeapSpKAdd streams input columns from memory and writes the output column to memory without storing any intermediate matrices. 
Thus, the I/O cost of the heap algorithm is $O(\sum_{i=1}^k{\nnz(\mA_i)})$.
For ER matrices, the computational and I/O complexity of HeapSpKAdd become $O(ndk\lg k)$ and $O(ndk)$, respectively.

\subsubsection{SPASpKAdd: k-way addition using a SPA}
Algorithm~\ref{algo:SPA} describes the addition of columns using a SPA. 
We represent SPA with two dense arrays \id{SPA} of length $m$ (number of rows in matrices) and \id{idx} of length $\nnz(\mB(:,j))$.
The $r$th entry of \id{SPA} stores the value at the $r$th row of the output $\nnz(\mB(r,j))$.
The \id{idx} array stores the indices of \id{SPA} with valid entries. 
Each iteration of the loop at line 4 in Algorithm~\ref{algo:SPA} processes the input column from the $i$ matrix. Consider an entry $(r,v)$ from the current column $\mA_i(:,j)$. If this row exists in the \id(idx) array (i.e., a valid entry exists at \id{SPA[r]}), the value $v$ is added to the corresponding location at \id{SPA} (line 6).
Otherwise, the row index $r$ is inserted into the \id{idx} array.
If sorted output is desired the \id{idx} array is sorted before generating the result. 
See Fig.~\ref{fig:AllAlgorithms}(e) for an example. 

{\bf Complexity.}
Since every nonzero entry from the input is inserted once to SPA, the computational complexity of the SPA based algorithm is $O(\sum_{i=1}^k{\nnz(\mA_i)})$.
The I/O complexity is also $O(\sum_{i=1}^k{\nnz(\mA_i)})$.

\subsubsection{HashSpKAdd: using k-way additions using a hash table}
Algorithm~\ref{algo:hash} shows the HashAdd operation that uses a hash table \id{HT} to add $j$th columns of all input matrices. 
The size of \id{HT} is a power of two and it is greater than $\nnz(B(:,j))$ (which is obtained via a symbolic phase discussed in Section~\ref{Sec:symbolic}).
The hash table \id{HT} stores $r,v$ tuples of row indices and values for all unique row indices in input columns. \id{HT} is initialized with $(-1,0)$ where a negative row index denotes an unused entry in the hash table.

Each iteration of the loop at line 3 in Algorithm~\ref{algo:hash} processes the input column from the $i$ matrix. Consider an entry $(r,v)$ from the current column $\mA_i(:,j)$.
A suitable hash function is used to map the row index $r$ to a hash index $h$ (line 5).
In our implementation, we used a multiplicative masking scheme $\Call{Hash}{r}= (a*r) \& (2^q-1)$, where $r$ is the row index used as the hash key, $a$ is a prime number, $2^q$ is the size of the hash table with $2^q\geq \nnz(B(:,j))$.
The bitwise and operator ($\&$) masks off the bottom $q$ bits so that the output of the hash function can be used as an index into a table of size $2^q$.
The while loop at line 6 search for the row index $r$ in \id{HT}. If such an entry is found, its value is updated in line 8. If not found (meaning that we encounter $r$ for the first time), the current tuple $(r,v)$ is stored in \id{HT} (line 7).
When a valid entry with a different row index is encountered (line 11), it indicates a collision in hash tables.
We used the linear probing scheme to resolve collisions.  
After we process all input columns, we insert all valid entries from the hash table to form the output. 
See Fig.~\ref{fig:AllAlgorithms}(f) for an example. 

{\bf Complexity.} 
Inserting a value into a Hash table takes, on the average case, $O(1)$ time.
Since every nonzero entry from the input is inserted to the hash table for processing, the average computational complexity of the hash based algorithm is $O(\sum_{i=1}^k{\nnz(\mA_i)})$.
The I/O cost of the hash algorithm is $O(\sum_{i=1}^k{\nnz(\mA_i)})$.
For ER matrices, both the computational and I/O complexity of HashSpKAdd become $O(ndk)$.

\begin{algorithm}[!t]
\caption{Computing $\nnz(B(:,j))$ using a hash table. 
}

\begin{algorithmic}[1]
\Procedure{HashSymbolic}{$\mA_1(:,j),...,\mA_k(:,j)$}
    \State $\id{HT}  \gets $ an array whose size is a power of two and greater than $\sum_{i=1}^k{\nnz(A_i)}$; initialized with $-1$
    \State nz $\gets $ 0
    \For{$i \gets 1$ to $k$}
        \For{$(r,v) \in \mA_i(:,j)$}
            \State $h\gets \Call{Hash}{r}$ \Comment{hash index from row index}
            \While {true}
            \If{$\id{HT[h]}$ is empty}
             \State nz $\gets $ nz + 1
             \State $\id{HT}[h]\gets  r$; {\bf break}
            \ElsIf{$r$ is found at $\id{HT[h]}$ }
             {\bf break}
            \Else \ $h\gets $ next hash index
            \EndIf
        \EndWhile
        \EndFor
    \EndFor
\State \Return nz
\EndProcedure
\end{algorithmic}
\label{algo:hashsymbolic}
\end{algorithm}

\subsection{The symbolic step to compute the output size}
\label{Sec:symbolic}
All k-way \spkadd algorithms need to know $\nnz(\mB(:,j))$ for every column of the output to pre-allocate necessary memory for the output. 
This estimation is also needed to determine the hash table size. 
We pre-compute $\nnz(\mB(:,j))$ using a symbolic phase as shown in Algorithm~\ref{algo:hashsymbolic}. Here we use hash based symbolic phase, but heap and SPA could also be used to compute $\nnz(\mB(:,j))$. 
The hash table in the symbolic phase needs to store indices only. 
Algorithm~\ref{algo:hashsymbolic} first creates a hash table of size $\sum_{i=1}^k\nnz(\mA_i)$ and initializes it with -1. 
Then, similar to the actual addition (Algorithm~\ref{algo:hash}), we search for the current row index $r$ in the hash table. The first time $r$ is found in the table, we increment the nonzero counter by one (lines 8-9). 
The computation and I/O complexity of Algorithm~\ref{algo:hashsymbolic} is similar to that of Algorithm~\ref{algo:hash}. 



\section{Parallel and Cache Efficient \spkadd Algorithms}
\subsection{Parallel algorithms}
The way we design 2-way and k-way \spkadd algorithms makes it very easy to parallelize them. 
Since each column of $\mB$ can be computed independently, we can parallelize line 4 of Algorithm~\ref{algo:twoway} and line 2 of Algorithm~\ref{algo:generickway}. 
This parallelization strategy does not require any thread synchronization and it does not depend on the data structure (heap, hash, SPA, etc.) used to add the $j$th columns of input matrices. 
In this case, we always execute Algorithms ~\ref{algo:heap}, \ref{algo:SPA} , \ref{algo:hash}, and \ref{algo:hashsymbolic} sequentially by a single thread. 
Note that parallelizing \spkadd as a single multiway merging problem (e.g., when each matrix is stored as a list of (rowidx, colidx, val) tuples in the coordinate format) is not a trivial task as it requires the lists to be partitioned among threads~\cite{3dspgemmsisc16}. 
Even though \spkadd (when dividing columns among threads) is an embarrassingly parallel problem, the choices of data structures, matrix shapes and sparsity patterns, and the memory subsystem of the computing platform play significant roles on the attained performance of \spkadd. 

{\bf The impact of data structures on the parallel performance.}
Our parallel \spkadd algorithms maintain thread-private data structures (because Algorithms ~\ref{algo:heap}, \ref{algo:SPA} , \ref{algo:hash}, and \ref{algo:hashsymbolic} are executed sequentially).
Hence the total memory requirements for different data structures across $T$ threads are as follows:
heap: $O(T\cdot k)$, SPA: $O(T\cdot m)$, Hash: $O(T\cdot\max_{i=1}^n{\nnz(\mB_i}))$.
The higher memory requirement of SPA may make the corresponding algorithm very slow because the algorithm accesses SPA at random locations. For example, if matrices have 32M rows, and 48 threads are used, the memory requirement for SPA could be more than 100GB.
In contrast the memory requirement of heap is very small. However, the heap algorithm accesses all matrices concurrently (see Figure~\ref{fig:AllAlgorithms}(d)). This could create bandwidth bottleneck if different matrices are stored in multiple sockets of a shared-memory platform.

{\bf Impact of matrix shapes and sparsity patterns
on load imbalance.}
The computations in different columns of \spkadd are fairly load balanced when nonzeros in input matrices are uniformly distributed such as in \erdosrenyi~matrices. 
However, for matrices with skewed nonzero distributions such as RMAT matrices, the computations can vary dramatically across columns. 
In the latter case, a static scheduling of threads hurts the parallel performance of the algorithm. 
In the symbolic phase we use total input non-zeros per column and in addition phase we use total output non-zeros per column to balance loads dynamically.

\begin{algorithm}[!t]
\caption{Computing $\nnz(B(:,j))$ using a sliding hash table. $T$: number of threads; $M$: total last-level cache; $m$: number of rows in matrices; $b$: number bytes needed for each entry in the hash table (4 bytes for the symbolic phase).
}

\begin{algorithmic}[1]
\Procedure{SlHashSymbolic}{$\mA_1(:,j),...,\mA_k(:,j)$}
    \State $\id{inz} \gets \sum_{i=1}^k{\nnz(A_i)}$
    \State $\id{parts} \gets \lceil(\id{inz} * b*T)/M\rceil$
    \State \id{nz} $\gets 0$ 
    \If{\id{parts}=1} \Comment{normal hash}
        \State \Call{HashSymbolic}{$\mA_1(:,j),...,\mA_k(:,j)$}
    \Else \Comment{sliding hash}
        \For{$i \gets 1$ to \id{parts}}
        \State $r_1\gets i*m/\id{parts}$; \ \ 
        $r_2\gets (i+1)*m/\id{parts}$
        \State \id{nz} $\gets $\id{nz} + \Call{HashSymbolic}{$\mA_1(r_1:r_2,j),...,\mA_k(r_1:r_2,j)$}    
        \EndFor
    \EndIf
    
\State \Return nz
\EndProcedure
\end{algorithmic}
\label{algo:slidinghashsymbolic}
\end{algorithm}

\subsection{The Sliding Hash algorithm} 
In Algorithm~\ref{algo:hashsymbolic} and Algorithm~\ref{algo:hash} we used a single hash table for the computation of the entire column of the output. These approaches work perfectly well as long as all hash tables used by all threads fit in the last level cache. 
To understand the memory requirements of hash tables, let $M$ be the size of last level cache in bytes, $T$ be the number of threads and $b$ be the number of bytes needed to store each entry in the hash table. When 32 bit indices and single-precision floating point values are stored, $b$ is 4 bytes for the symbolic phase and 8 bytes for the addition phase. 
Then, the total hash table memory required for the symbolic phase is MemSym = $b*T*\sum_{i=1}^k{\nnz(\mA_i(:,j))}$ bytes. 
Similarly, the total hash table memory required for addition phase is MemAdd= $b*T*\nnz(\mB(:,j))$ bytes. 
If either MemSym or MemAdd is greater than $M$, the hash table goes out of cache. Since hash tables are accessed randomly, out of cache accesses are quite expensive. 

\begin{algorithm}[!t]
\caption{SpKAdd using a sliding hash table. $T$: number of threads; $M$: total last-level cache; $m$: number of rows in matrices; $b$: number bytes needed for each entry in the hash table (8 bytes for 32 bit indices and values).
}

\begin{algorithmic}[1]
\Procedure{SlHashAdd}{$\mA_1(:,j),...,\mA_k(:,j)$}
    \State $\id{onz} \gets \Call{SlHashSymbolic}{\mA_1(:,j),...,\mA_k(:,j)}$
    \State $\id{parts} \gets \lceil(\id{onz} * b*T)/M\rceil$
    \State \id{nz} $\gets 0$ 
    \If{\id{parts}=1} \Comment{normal hash}
        \State \Call{HashAdd}{$\mA_1(:,j),...,\mA_k(:,j)$}
    \Else \Comment{sliding hash}
        \For{$i \gets 1$ to \id{parts}}
        \State $r_1\gets i*m/\id{parts}$ ; \ \ 
        $r_2\gets (i+1)*m/\id{parts}$
        \State $\mB(r_1{:}r_2,j){\gets}$  \Call{HashAdd}{$\mA_1(r_1:r_2,j), ..., \mA_k(r_1:r_2,j)$}    
        \EndFor
    \EndIf
    
\State \Return $\mB(:,j)$
\EndProcedure
\end{algorithmic}
\label{algo:slidinghash}
\end{algorithm}

\begin{table*}[!t]
    \centering
     \caption{Summary of all \spkadd algorithms when adding $k$ ER matrices each with $d$ nonzeros per column on average. $T$ denotes the number of threads in parallel algorithms and $M$ denotes the size of the last-level cache. The corresponding complexities for general matrices are shown in the main text.}
    \vspace{-5pt}
    \begin{tabular}{c l  c c c  c c }
    \toprule
      Algorithm  & Work & I/O (from memory) & Data structure memory & Data accesses & Need sorted inputs?\\
      \toprule
     2-way Incremental & $O(k^2nd)$  & $O(k^2nd)$ & - & streamed & yes\\ 
     2-way Tree & $O(knd \lg k)$ & $O(knd \lg k)$ & - & streamed & yes\\ 
     k-way Heap & $O(knd \lg k)$ &  $O(knd)$ & $O(Tk)$& concurrent streaming of $k$ columns & yes\\ 
     k-way SPA & $O(knd)$ &  $O(knd)$ & $O(Tm)$ & random access SPA & no\\ 
     k-way Hash & $O(knd)$  & $O(knd)$ & $O(Tkd)$ & random access hash table & no\\ 
     k-way Sliding Hash & $O(knd)$ &  $O(knd)$ & $O(M)$  &  in-cache hash table & no\\ 
     \bottomrule
    \end{tabular}
   
    \label{tab:summary}
\end{table*}

To address this problem, we designed a modified hash algorithm that limit the size of hash tables so that they fit within the last level cache. For example, when MemAdd is greater than $M$, we create each hash table with no more than  $M/(b*T)$ entries so that all hash tables across threads fit in $M$ bytes. Since such a collection of hash tables cannot cover the entire output column, we slide a hash table along the column to produce outputs. Hence, we call it the sliding hash algorithm. 

Algorithm~\ref{algo:slidinghashsymbolic} and Algorithm~\ref{algo:slidinghash} describes the symbolic and \spkadd algorithms using sliding hash tables. 
In Algorithm~\ref{algo:slidinghashsymbolic} we divide the required memory by $M$ to determine the number of parts (line 3). If the number of parts is one (i.e., all hash tables fit in $M$ byes), we simply call Algorithm~\ref{algo:hashsymbolic} (lines 5-6). If more parts are needed, we partition rows equally (using binary searches) and then call Algorithm~\ref{algo:hashsymbolic} for each part (lines 8-11).
Similarly, Algorithm~\ref{algo:slidinghashsymbolic} identifies the memory requirement for all hash tables and uses the sliding hash strategy (lines 8-11) when needed. 
Thus, the sliding hash algorithms ensure that hash tables accesses are always performed within the cache, which gives significant performance boosts when the number of input matrices is large or when input matrices are denser.
Note that hash tables in the symbolic phase could be larger than hash tables in the addition phase. For example, when $cf$ of an \spkadd operation is 10, the symbolic phase may demand $10\times$ more memory for hash tables. Hence, the sliding hash table is more beneficial for the symbolic phase.

\vspace{-3pt}
\subsection{Summary of all \spkadd algorithms}
\vspace{-3pt}
Table~\ref{tab:summary} shows the complexity summary of algorithms when adding $k$ ER matrices each with $d$ nonzeros per column on average.
We observe that 2-way and heap algorithms are not work efficient, and 2-way algorithms access more data from memory than other algorithms. 
Thus, we expect 2-way and heap algorithms to run slower than other algorithms. 
Our experimental results confirm this theoretical prediction. 
The memory requirements for various data structures also vary significantly. 
SPA requires the most memory, while the heap algorithm needs the least memory (assuming that $k\ll m$). 
The sliding hash algorithm allocates hash tables based on the available cache. 
Both SPA and hash algorithms accesses their data structures randomly. Random accesses can be a potential performance bottleneck if hash tables and SPA do not fit in the cache. The sliding hash table mitigates this problem by limiting the size of hash table. Finally, SPA and hash algorithms can operate with unsorted input columns, but 2-way and heap algorithms require sorted inputs.  







\begin{table}[!t]
\centering
\caption{Overview of the evaluation platforms. }
\vspace{-3pt}
\label{tab:hardware}
\begin{tabular}{c|c|c|c|c|}
\cline{2-5}
\textbf{}                                          & \textbf{Property} & \textbf{\begin{tabular}[c]{@{}c@{}}Intel\\ Skylake  8160\end{tabular}}   & \textbf{\begin{tabular}[c]{@{}c@{}}AMD\\ EPYC 7551\end{tabular}} & \textbf{\begin{tabular}[c]{@{}c@{}} Cori Node\\ Intel KNL\end{tabular}} \\ \hline
\multicolumn{1}{|c|}{\multirow{4}{*}{\parbox[t]{2mm}{\multirow{1}{*}{\rotatebox[origin=c]{90}{Core}}}}}        & Clock             & 2.10 GHz                                                           & 2 GHz                                                            & 1.4 GHz                                                         \\[0.18ex] 
\multicolumn{1}{|c|}{}                             & L1 cache          & 32KB   &                              32KB                        & 32KB                                                              \\[0.18ex] 
\multicolumn{1}{|c|}{}                             & L2 cache          & 1MB   &                              512KB                        & $\times$ \\[0.18ex] 
\multicolumn{1}{|c|}{}                             & LLC          & 32MB                                                                & 8MB                                                            & 34MB                                                                \\[0.18ex] \hline
\multicolumn{1}{|c|}{\multirow{3}{*}{\parbox[t]{2mm}{\multirow{1}{*}{\rotatebox[origin=c]{90}{\vspace{-2cm} Node}}}}} & Sockets           & 2                                                                  & 2                                                                & 1                                                               \\ [0.18ex]
\multicolumn{1}{|c|}{}                             & Cores/soc.        & 24                                                                 & 32                                                    & 68                                                              \\ [0.18ex]
\multicolumn{1}{|c|}{}                             & Memory            & 256GB                                                              & 128GB                                                            & 108GB                                                            \\ \hline
\end{tabular}
\end{table}

\begin{figure*}[!t]
    \centering
    \includegraphics[width=0.95\linewidth]{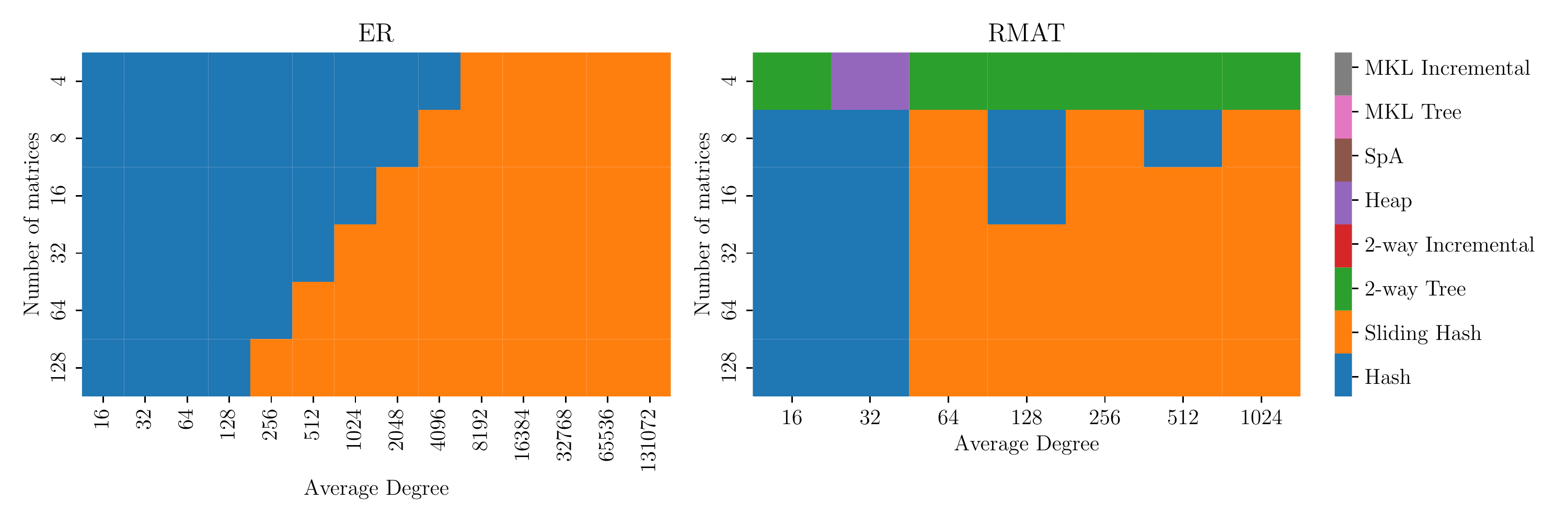}
    \vspace{-7pt}
    \caption{Best performing algorithms with respect to the number of matrices ($k$) and their sparsity ($d$). Experiments were run on the Intel Skylake processor. \vspace{-5pt}}
    \label{fig:benchmarking}
\end{figure*}

\section{Results}
\vspace{-3pt}
\subsection{Experimental setup}
\label{subsec:setup}
\vspace{-3pt}
\textbf{Experimental platforms.}
Most of our experiments are conducted on 
two different servers with Intel and AMD processors as described in Table \ref{tab:hardware}. 
To demonstrate the impact of our \spkadd algorithms on a distributed SpGEMM algorithm, we ran some experiments on the KNL partition of Cori supercomputer at NERSC. 
We have implemented our in C/C++ programming language with OpenMP multi-threading.
For distributed memory experiments, we plugged in our code in the sparse SUMMA algorithm available in the Combinatorial BLAS (CombBLAS) library~\cite{combblas2}. 

{\bf Datasets.}
We use two types of matrices for the evaluation. 
We generate synthetic matrices using matrix generator, and also use publicly available protein similarity matrices.
We use R-MAT~\cite{chakrabarti2004r}, the recursive matrix generator, to generate
two different non-zero patterns of synthetic matrices represented
as ER and RMAT. ER matrix represents \erdosrenyi~ random graphs,
and RMAT represents graphs with power-law degree distributions
used for Graph500 benchmark. These matrices are generated with
R-MAT seed parameters; $a {=} b {=} c {=} d {=} 0.25$ for ER matrix and
$a {=} 0.57,b {=} c {=} 0.19,d {=} 0.05$ for G500 matrix. 
Most of our experiments use rectangular matrices with more rows than columns ($m>n$).
To generate $k$ RMAT matrices for \spkadd experiments, we create a matrix $m\times n$ matrix and then split this matrix along the column to create $k$  $m\times m/k$ matrices.

We also use  protein-similarity  networks Eukarya (3M rows, 3M columns, 360M nonzeros) and Isolates (35M rows, 35M columns, 17B nonzeros) that represent networks  generated  from isolate genomes in the IMG database and are publicly available  with  the HipMCL software~\cite{Azad2018}.  We also used Metaclust50 (282M rows, 282M columns, 37B nonzeros)  that stores  similarities  of proteins  in  Metaclust50  (https://metaclust.mmseqs.com/) dataset which contains predicted genes from metagenomes and metatranscriptomes of assembled contigs from IMG/Mand NCBI. 
We used these matrices to demonstrate the use of \spkadd in distributed SpGEMM algorithms.

\begin{table*}[!t]
    \centering
    \caption{Runtime (sec) of different algorithms for different values of $k$ and different average nonzeros per column ($d$) of ER matrices on the Intel Skylake processor (48 cores).  Green cells represent the smallest runtime in each column. }
    \vspace{-5pt}
    \label{tab:detailER}
    \begin{tabular}{|c| l | l | l | l | l | l | l | l | l |}
    \hline
    \multirow{2}{*}{Algorithm}
        & \multicolumn{3}{c|}{$d=16$} & \multicolumn{3}{c|}{$d=1024$} & \multicolumn{3}{c|}{$d=8192$} \\
        \cline{2-10}
        & \multicolumn{1}{c|}{$k=4$} & \multicolumn{1}{c|}{$k=32$} & \multicolumn{1}{c|}{$k=128$} & \multicolumn{1}{c|}{$k=4$} & \multicolumn{1}{c|}{$k=32$} & \multicolumn{1}{c|}{$k=128$} & \multicolumn{1}{c|}{$k=4$} & \multicolumn{1}{c|}{$k=32$} & \multicolumn{1}{c|}{$k=128$} \\
        \hline
    2-way Incremental & 0.0022 & 0.0316 & 0.4506 & 0.0357 & 0.6618 & 5.7806 & 0.1746 & 2.7200 & could not run \\
    MKL Incremental & 0.0588 & 0.6924 & 3.9971 & 0.1638 & 3.4144 & 29.1978 & 0.4717 & 12.9322 & 172.2600 \\
    2-way Tree & 0.0019 & 0.0186 & 0.0832 & 0.0279 & 0.2440 & 1.2798 & 0.4416 & 0.8273 & 4.0545 \\
    MKL Tree & 0.0499 & 0.5237 & 2.2155 & 0.1425 & 1.8711 & 8.2814 & 0.4121 & 5.4968 & 26.1705  \\
    Heap & 0.0037 & 0.0112 & 0.0374 & 0.0730 & 0.5570 & 2.1732 & 0.5914 & 4.5293 & 16.1049 \\
    SPA & 0.1237 & 0.1274 & 0.1309 & 0.1351 & 0.2844 & 0.8173 & 0.2579 & 1.3776 & 4.5269  \\
    Hash & \cellcolor{green}0.0007 & \cellcolor{green}0.0016 & \cellcolor{green}0.0083 & \cellcolor{green}0.0110 & 0.1362 & 0.4463 & 0.1312 & 0.8062 & 4.3987 \\
    Sliding Hash & 0.0021 & 0.0045 & 0.0162 & 0.0272 
    & \cellcolor{green}0.0936 & \cellcolor{green}0.3330 & \cellcolor{green}0.0993 & \cellcolor{green}0.5773 & \cellcolor{green}1.8096 \\
    \arrayrulecolor{black}\hline
    \end{tabular}
\end{table*}

\begin{table*}[!t]
    \centering
    \caption{Runtime (sec) of different algorithms for different values of $k$ and different average nonzeros per column ($d$) of RMAT matrices on the Intel Skylake processor (48 cores). Green cells represent the smallest runtime in each column.}
    \vspace{-5pt}
    \label{tab:detailRMAT}
    \begin{tabular}{|c| l | l | l | l | l | l | l | l | l |}
    \hline
    \multirow{2}{*}{Algorithm}
        & \multicolumn{3}{c|}{$d = 16$} & \multicolumn{3}{c|}{$d = 64$} & \multicolumn{3}{c|}{$d = 512$} \\
        \cline{2-10}
        & \multicolumn{1}{c|}{$k=4$} & \multicolumn{1}{c|}{$k=32$} & \multicolumn{1}{c|}{$k=128$} & \multicolumn{1}{c|}{$k=4$} & \multicolumn{1}{c|}{$k=32$} & \multicolumn{1}{c|}{$k=128$} & \multicolumn{1}{c|}{$k=4$} & \multicolumn{1}{c|}{$k=32$} & \multicolumn{1}{c|}{$k=128$} \\
        \hline
    2-way Incremental & 0.1155 & 0.8078 & 3.1727 & 0.3269 & 2.4913 & 9.3310 & 2.1307 & could not run & could not run \\
    MKL Incremental & 0.9839 & 6.0211 & 17.1921 & 3.5291 & 18.3133 & 47.9294 & 24.6052 & 109.5036 & 264.0963 \\
    2-way Tree & 0.0774 & 0.2315 & 0.4691 & 0.2358 & 0.5673 & 0.9890 & 1.5213 & 3.3897 & 4.6937 \\
    MKL Tree & 0.8670 & 2.2720 & 4.1899 & 2.9447 & 5.9594 & 8.3661 & 19.8781 & 31.0663 & 33.3237 \\
    Heap & 0.1371 & 0.1384 & 0.1960 & 0.2422 & 1.0980 & 0.7603 & 1.8223 & 3.5277 & 6.7888 \\
    SPA & 0.2467 & 0.2448 & 0.2386 & 0.5705 & 0.5659 & 0.6199 & 4.3047 & 4.6626 & 9.2565 \\
    Hash & \cellcolor{green}0.1068 & \cellcolor{green}0.0739 & \cellcolor{green}0.0719 & 0.3240 & 0.3121 & 0.2925 & \cellcolor{green}1.7651 & 1.8187 & 1.8500 \\
    Sliding Hash & 0.1251 & 0.0794 & 0.0762 & \cellcolor{green}0.3206 & \cellcolor{green}0.2625 & \cellcolor{green}0.2562 & 1.7909 & \cellcolor{green}1.7053 & \cellcolor{green}1.5471 \\
    \arrayrulecolor{black}\hline
    \end{tabular}
\end{table*}

\vspace{-3pt}
\subsection{Summary of performance with respect to sparsity patterns}
\vspace{-3pt}
At first, we study the best performing algorithms for different settings of input matrices including the number of matrices ($k$), the average number of nonzeros ($d$) in each column of a matrix, and their sparsity patterns (uniformly distribution in ER and scale-free distribution in RMAT).

The left subfigure in Fig.~\ref{fig:benchmarking} shows that hash and sliding hash based \spkadd algorithms are always the best performers for ER matrices. When the number of matrices is small or when the matrices are sparser, the hash tables from all threads fit within the cache. In these settings, there is no benefit of using the sliding hash algorithm. When the hash tables start to spill out from cache, the benefit of sliding hash tables becomes visible. 
The boundary between blue and orange regions is mostly determined by the last level cache size and the number threads sharing the cache. 
For example, when adding $k=128$ matrices with average degree of $d=512$, the expected number of nonzeros in an output column is 512*128=65,536.
If each hash table entry needs 12 bytes to store  (rowid, val) tuple and there are 48 threads, the total memory requirement to store all hash tables is about 38MB. 
Since our Intel Skylake processor has 32MB last level cache, the sliding hash is faster for $k=128, d=512$. No other algorithms perform better than the hash \spkadd for ER matrices. 

The right subfigure in Fig.~\ref{fig:benchmarking} shows that hash and sliding hash based \spkadd algorithms are the best performers for RMAT matrices when $k\geq 8$. 
For $k=4$, heap or 2-way tree algorithms perform the best. This is because the addition of RMAT matrices are dominated by denser columns. If we have just $k=4$ columns to add and one of them is dense, 2-way addition algorithms can simply stream the denser column and add other sparse columns by utilising  thread private L1 cache.  With more matrices, the \spkadd algorithms may need to add multiple denser columns in an iteration, which makes the hash algorithm perform better than its peers. 
Similar to the ER matrices, the performance of hash and sliding hash is determined by the size of the hash table. Sliding hash helps lower the runtime when the input matrices are denser and hash tables tend to go out of the cache.

\begin{figure*}[t!]
    \centering
    \includegraphics[width=0.95\linewidth]{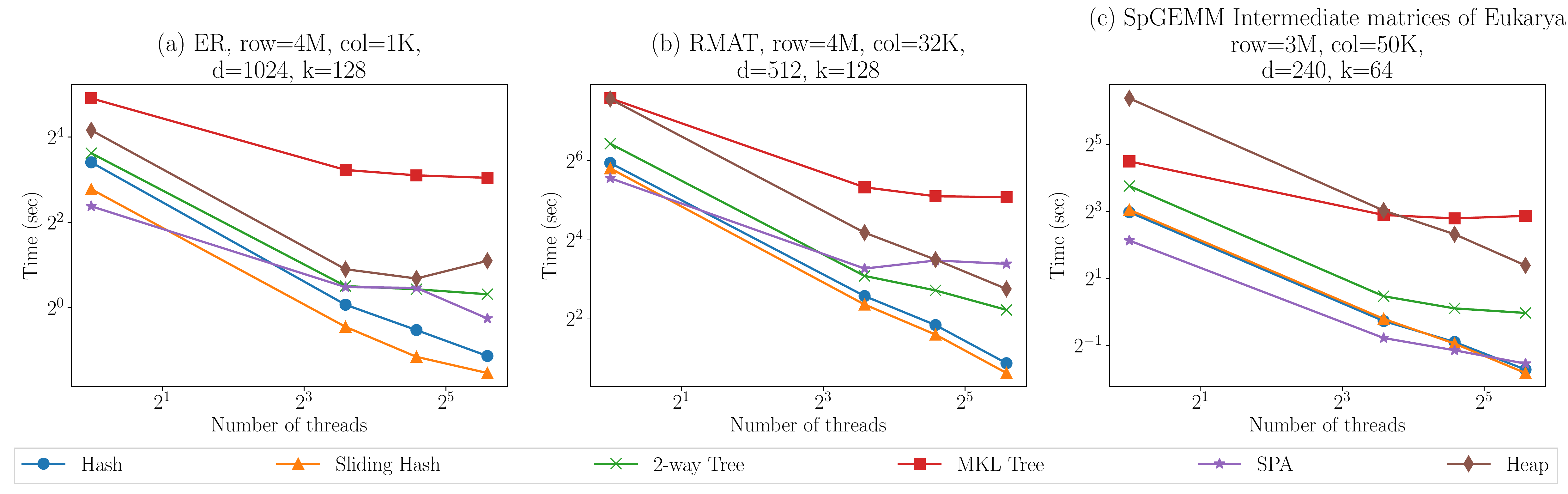}
    \vspace{-5pt}
    \caption{Strong scaling of various \spkadd algorithms on the Intel processor. \vspace{-5pt}}
    \label{fig:scaling}
\end{figure*}

While Fig.~\ref{fig:benchmarking} provides an overall summary of the best performing algorithms, it does not show their actual runtimes.  For a representative subset of matrices, we show the runtime of various algorithms in  Table~\ref{tab:detailER} and Table~\ref{tab:detailRMAT}.
These tables confirm the performance observed in Fig.~\ref{fig:benchmarking}. 
Additionally, we note few properties of various algorithms observed in Table~\ref{tab:detailER} and Table~\ref{tab:detailRMAT}. 
(a) {\bf The performance of the 2-way tree algorithm improves with matrix density}. With a fixed $k=128$ for ER matrices in Table~\ref{tab:detailER}, the performance of 2-way tree relative to hash is: $10\times$ slower when $d=16$, $3\times$ slower when $d=1024$, and $1.1\times$ faster when $d=8196$. 
Similarly, for RMAT matrices with a fixed $k=128$ in Table~\ref{tab:detailRMAT}, the performance of 2-way tree relative to hash is: $6\times$ slower when $d=16$, $3\times$ slower when $d=1024$, and $2\times$ slower when $d=8196$.
This is expected as hash tables become larger for denser matrices, making the 2-way tree algorithm more efficient. However, the sliding hash algorithm is still the fastest at the extreme setting with $d=8196$.
(b) {\bf The SPA \spkadd algorithm is as efficient as the hash \spkadd algorithm for denser matrices} as seen in the last column of Table~\ref{tab:detailER}. This is expected because the hash table size ($\nnz(B(:,j))$) becomes $O(m)$ for dense output columns. This also suggests that the benefits of sliding hash can also be observed in the SPA algorithm if we partition the SPA array based on row indices~\cite{patwary2015parallel}.
(c) {\bf 2-way \spkadd using MKL is significantly slower than other algorithms.} The MKL-based 2-way \spkadd algorithms do not show promising results in our experiments. This especially true for larger values of $k$. 
\vspace{-3pt}
\subsection{Scalability}
\vspace{-3pt}
We conducted the strong scaling experiments on the Intel Skylake processors with up to 48 threads (at most one thread per core). 
Fig.~\ref{fig:scaling} shows the scalability results with (a) ER matrices, (b) RMAT matrices, and (c) 64 lower rank matrices generated when running a distributed SpGEMM algorithm with the Eukarya matrix. 
We observe that most algorithms scale almost linearly except 2-way \spkadd algorithms (MKL Tree and 2-way tree) and SPA-based \spkadd.
Our load balancing technique ensures that k-way \spkadd implementations scale well for matrices with skewed nonzero distributions.  
The performance of SPA suffers at high thread counts because of $O(Tm)$ memory requirement where $T$ is the number of threads and $m$ is the number of rows in matrices. 
However, the sequential SPA-SpKAdd performs better than other algorithms on a single thread because a SPA with 4M entries can still fit in the cache that is not shared among threads. However, for matrices with more rows (e.g., more that 32M rows), SPA would go out of cache even with a single thread. 

\begin{figure*}[!t]
    \centering
    \includegraphics[width=0.9\textwidth]{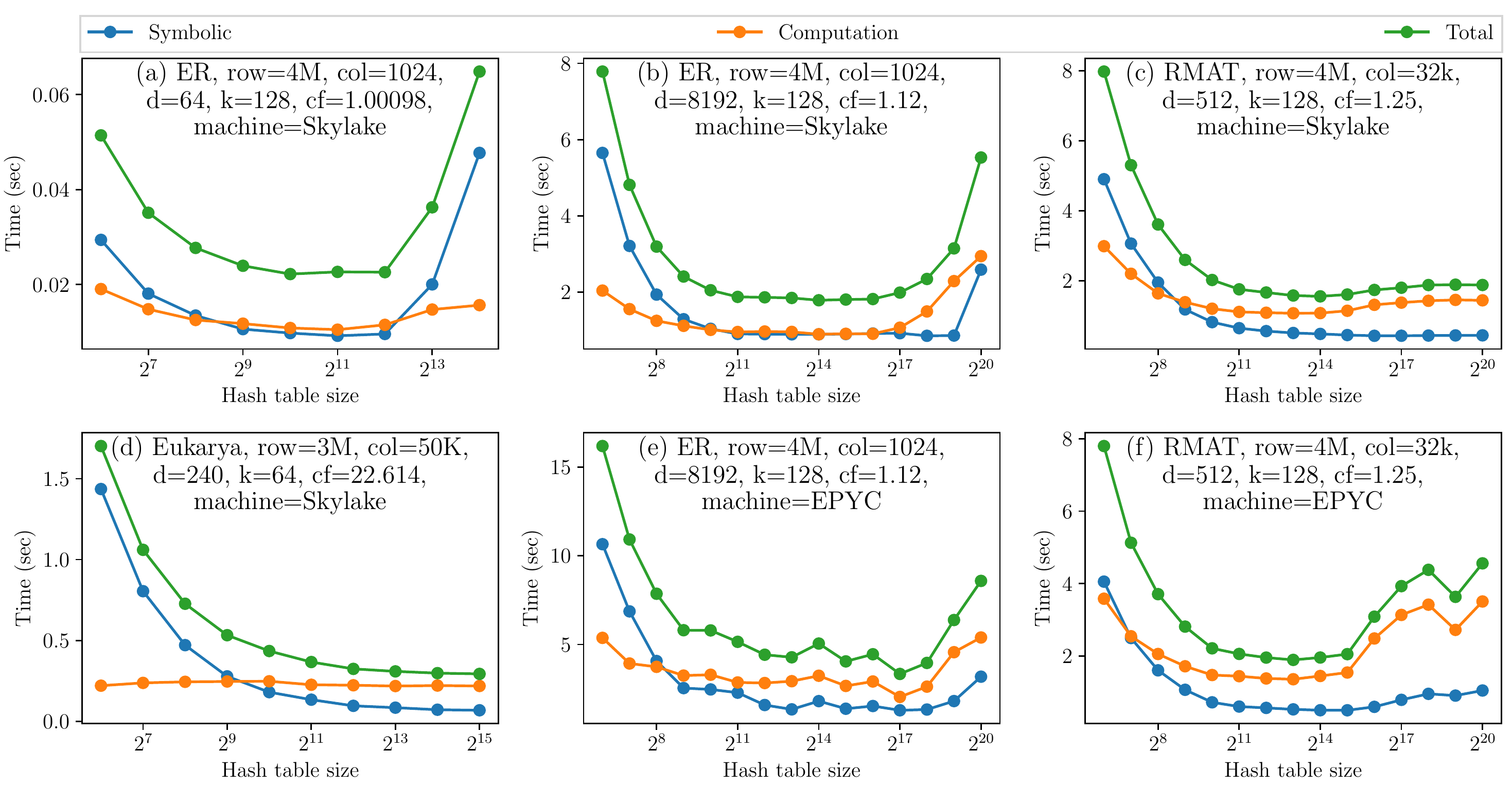}
    \vspace{-5pt}
    \caption{Determining optimum hash table size for the sliding hash algorithm. (a), (b), (c) and (d) are on the Intel Skylake processor. (e) and (f) are on AMD EPYC. The number of partitions (line 3 in Algorithm~\ref{algo:slidinghash} and ~\ref{algo:slidinghashsymbolic}) depends on the selected hash table sizes. The rightmost points in all figures denote hash table sizes that need no partitioning of input columns. \vspace{-8pt}}
    \label{fig:window}
\end{figure*}

\subsection{Understanding the performance of the hash algorithms}
\vspace{-3pt}
Since hash and sliding hash algorithms are among the best performers, we aim to understand when and why they perform better than their peers. 
Fig.~\ref{fig:window} shows the runtime of the sliding hash algorithm for different sizes of hash tables. 
The number of partitions (line 3 in Algorithm~\ref{algo:slidinghash} and ~\ref{algo:slidinghashsymbolic}) depends on the selected hash table sizes. 
For example, in Fig.~\ref{fig:window}(b), the expected number of nonzeros across $k$ input columns is about 1M. 
Thus, a hash table with 16K entries will translates into 64 partitions, but a hash table with with 1M entries translates into one partition where no sliding hash is used. 

\begin{table}[!t]
    \centering
    \caption{Last level cache misses on Intel Skylake.}
    \vspace{-3pt}
\begin{tabular}{|c | l l|}
  \hline
  \multirow{2}{*}{Cases (Fig.~\ref{fig:window})} 
      
      & \multicolumn{2}{c|}{LL miss read}\\             
& Sliding Hash & Hash \\  \hline
  (a)  & 1.8M & 1.4M \\
  (b)   & 214M & 734M \\
  (c)  & 344M & 409M \\
  (d) & 150M & 152M \\
  \hline
\end{tabular}
\label{tab:cache-profile}
\vspace{-5pt}
\end{table}

The optimum hash table sizes observed in Fig.~\ref{fig:window} are related to the cache sizes in different processors.
Fig.~\ref{fig:window}(a) is a case with very small matrices where the L1 cache plays a major role. Since our Intel processor has 32KB L1 cache per core, the L1 cache can hold a hash table with 4K entries (assuming that 8 bytes are needed for each entry in the hash table).
That is why the optimum hash table size is observed near 4K in Fig.~\ref{fig:window}(a).
For bigger matrices, the last level L3 cache influence the optimum hash table size. 32MB L3 cache shared among threads can hold hash tables with 
64K entries. 
Hence, we observe the optimum hash table sizes near 64K in Fig.~\ref{fig:window}(b) and (c).
The AMD processor used in our experiment has a smaller L3 cache (8MB). 
As a result, the optimum hash table size is smaller Fig.~\ref{fig:window}(e) and (f) that their corresponding sizes in Fig.~\ref{fig:window}(b) and (c).
We experimentally validate the aforementioned relations between hash table and cache sizes by using the Cachegrind profiler.
Table \ref{tab:cache-profile} reports the number of LL cache misses for the hash and sliding hash algorithms with various input matrices shown in Fig.~\ref{fig:window}.
As expected, for cases (b) and (c), the sliding hash encounter fewer LL cache misses than the hash algorithm. 
However, the sliding hash shows no observable benefits for cases (a) and (d) because these cases do not require large hash tables. 
The compression factor \spkadd also plays a role in the relative runtimes of the symbolic and addition phases of {\tt spkadd}.
For example, the Eukarya addition in Fig.~\ref{fig:window}(d) has a $\id{cf}=22.6$. Hence, the symbolic phase needed hash tables that are $27\times$ larger than the hash tables needed in the addition phase. Hence, the symbolic phase is very sensitive to the hash table size in Fig.~\ref{fig:window}(d), but the addition phase remains insensitive.



\begin{figure*}[t!]
    \centering
    \includegraphics[width=.95\linewidth]{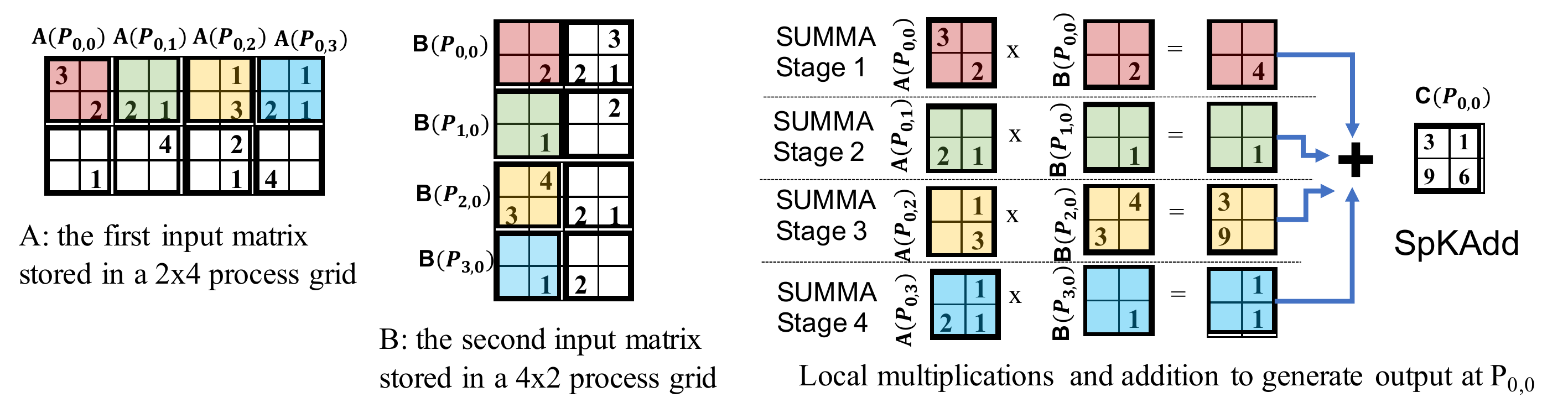}
    \vspace{-5pt}
    \caption{The use of \spkadd in distributed sparse SUMMA  algorithm (with stationary output). Here, the first input matrix is distributed on a $2\times 4$ process grid and the second matrix is stored on a $4\times 2$ grid. $\mA(P_{0,0})$ denotes the submatrix of $\mA$ stored in process $P_{0,0}$. Local submatrices in different processes are shown with thick grids. For this configuration, SUMMA operates in four phases where input submatrices are broadcasted along the process grid row and columns. We show the computation at $P_{0,0}$ where four local multiplications are performed, and the intermediate matrices are added using an \spkadd operation. \vspace{-15pt}}
    \label{fig:summa}
\end{figure*}

\begin{figure}[t!]
    \centering
    \includegraphics[width=0.95\linewidth]{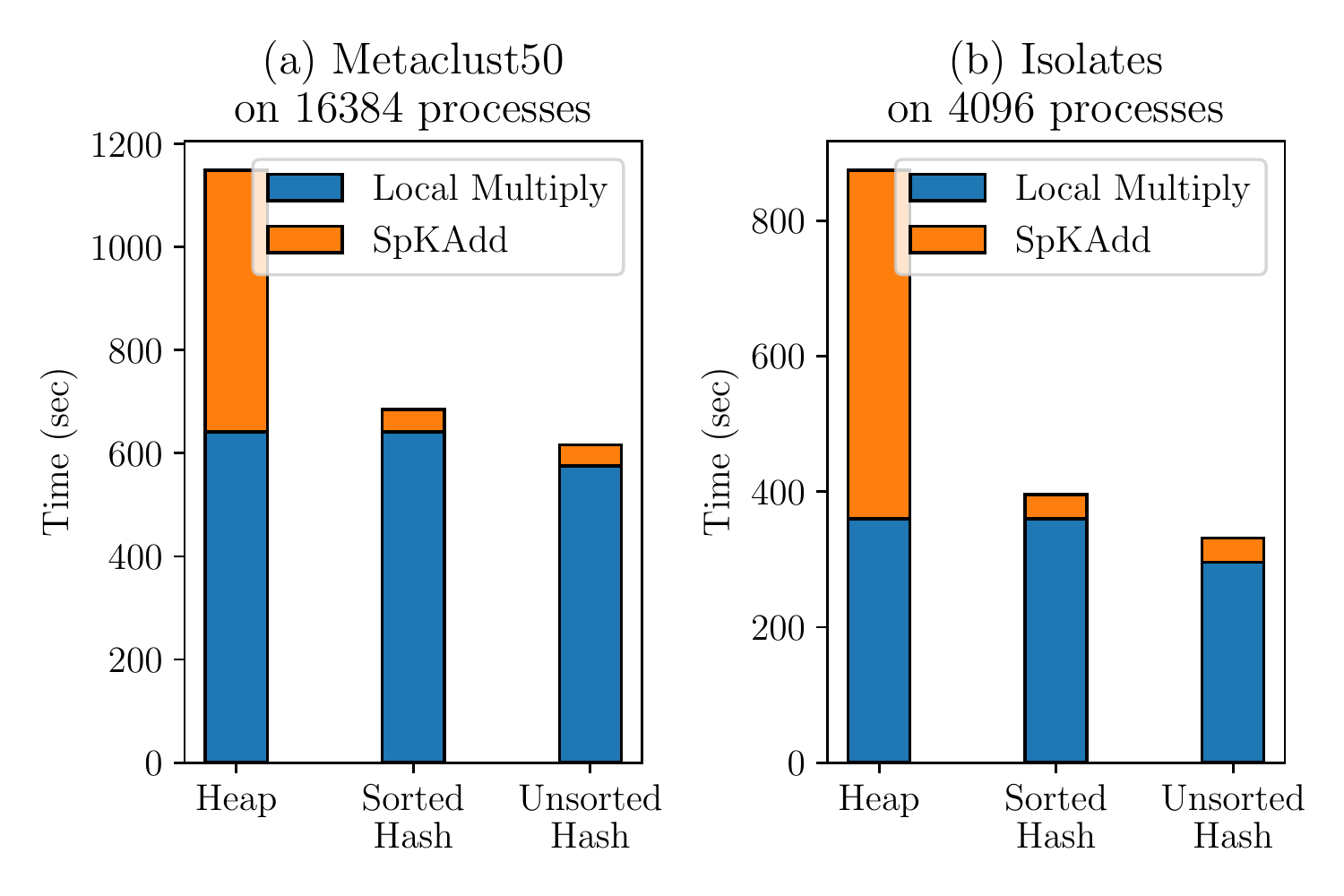}
    \vspace{-7pt}
    \caption{Effect of \spkadd in distributed memory SpGEMM running on Cori KNL nodes. In both cases, square process grid and 8 threads per process were used. \vspace{-5pt}}
    \label{fig:spgemm}
\end{figure}

 \subsection{Impact on the distributed SpGEMM algorithm}
 \vspace{-3pt}
 Among many applications of \spkadd, we discuss an application in distributed memory SpGEMM algorithms based on sparse SUMMA~\cite{Buluc2012}.
 Fig.~\ref{fig:summa} shows an example of sparse SUMMA with 8 processes. In this example, the first and second input matrices are distributed on a $2\times 4$ and $4\times 2$ logical process grids, respectively.
 After different stages of the algorithm, intermediate results from all stages are added to get the final results at different processes. 
 The last computational step of this sparse SUMMA algorithm needs an \spkadd operation as shown in Fig.~\ref{fig:summa} for process $P_{0,0}$. 
 
 To test the utility of our \spkadd algorithms, we plugged our hash implementation in the SpGEMM code inside the CombBLAS library~\cite{combblas2} and compare its performance with an existing heap \spkadd implementation.
 Since sparse SUMMA has two computational kernels (local SpGEMM and \spkadd), we show the runtime of both computational steps by excluding the communication costs. 
 Fig.~\ref{fig:spgemm} shows the result for two large scale distributed executions with 16384 processes for Metaclust50 and 4096 processes for Isolates-small respectively. 
 Fig.~\ref{fig:spgemm} shows that the  hash algorithms reduces the cost of \spkadd by an order of magnitude when compared with the existing heap algorithm. 
 Since hash \spkadd does not need sorted inputs, the preceding local multiplication can avoid sorting intermediate results. Skipping sorting in the local multiplications can make it 20\% faster as shown in Fig.~\ref{fig:spgemm}.
 
 \section{Conclusions}
 \vspace{-5pt}
 Despite being widely used in graph accumulation, matrix summarizing, and distributed-memory SpGEMM algorithms, \spkadd algorithms are not well studied in the literature.  
 In this paper, we theoretically and experimentally showed that \spkadd implemented using off-the-shelf 2-way additions is not work efficient and accesses more data from memory. 
We develop a family of shared-memory parallel \spkadd algorithms that rely on heap, SPA, and hash table data structures to add matrices column by column. Among these algorithms, a variant of hash algorithm is shown to be work efficient, uses spatial locality when accessing hash tables, and works with both sorted and unsorted input matrices. A limitation of the presented algorithms is our assumption that all input matrices fit in the main memory. When this assumption is not true (because the memory is limitated or matrices arrive in batches~\cite{selvitopi2020optimizing}), we can still arrange input matrices in multiple batches and then use \spkadd for each batch. 
We leave streaming \spkadd algorithms as a future work.

\bibliographystyle{IEEEtran}
\bibliography{Ref}

\end{document}